\documentclass[useAMS,usenatbib]{mn2e}

\usepackage{natbib}
\usepackage{graphicx}
\usepackage{amssymb}
\usepackage{color}
\usepackage{caption}

\title[Revisiting the `forbidden' region]{Revisiting the `forbidden' region: AGN radiative feedback with radiation trapping}
\author[ ]
{W. Ishibashi$^{1,}$$^{2}$\thanks{E-mail: wako@ast.cam.ac.uk}, A. C. Fabian$^{1}$, C. Ricci$^{3,}$$^{4,}$$^{5,}$$^{6}$ and A. Celotti$^{7,}$$^{8,}$$^{9}$
\footnotemark[0]\\
$^{1}$Institute of Astronomy, Madingley Road, Cambridge CB3 0HA \\
$^{2}$Physik-Institut, Universitat Zurich, Winterthurerstrasse 190, 8057 Zurich, Switzerland 
\footnotemark[0]\\
$^{3}$N\'ucleo de Astronom\'ia de la Facultad de Ingenier\'ia, Universidad Diego Portales, Av. Ej\'ercito Libertador 441, Santiago, Chile \\
$^{4}$Instituto de Astrof\'{\i}sica, Facultad de F\'{i}sica, Pontificia Universidad Cat\'{o}lica de Chile, Casilla 306, Santiago 22, Chile \\
$^{5}$Chinese Academy of Sciences South America Center for Astronomy, Camino El Observatorio 1515, Las Condes, Santiago, Chile \\
$^{6}$Kavli Institute for Astronomy and Astrophysics, Peking University, Beijing 100871, China \\
$^{7}$SISSA, via Bonomea 265, 34136 Trieste, Italy \\
$^{8}$INAF - Osservatorio Astronomico di Brera, via Bianchi 46, 323807 Merate, Italy \\
$^{9}$INFN - Sezione di Trieste, via Valerio 2, 34127 Trieste, Italy
}

\begin{document}

\pdfminorversion=4

\date{Accepted ? Received ?; in original form ? }

\pagerange{\pageref{firstpage}--\pageref{lastpage}} \pubyear{2012}

\maketitle

\label{firstpage}

\begin{abstract} 
Active galactic nucleus (AGN) feedback, driven by radiation pressure on dust, is an important mechanism for efficiently coupling the accreting black hole to the surrounding environment. Recent observations confirm that X-ray selected AGN samples respect the effective Eddington limit for dusty gas in the plane defined by the observed column density versus the Eddington ratio, the so-called $N_{\rm H} - \lambda$ plane. A `forbidden' region occurs in this plane, where obscuring clouds cannot be long-lived, due to the action of radiation pressure on dust. Here we compute the effective Eddington limit by explicitly taking into account the trapping of reprocessed radiation (which has been neglected in previous works), and investigate its impact on the $N_{\rm H} - \lambda$ plane. We show that the inclusion of radiation trapping leads to an enhanced forbidden region, such that even Compton-thick material can potentially be disrupted by sub-Eddington luminosities. We compare our model results to the most complete sample of local AGNs with measured X-ray properties, and find good agreement. Considering the anisotropic emission from the accretion disc, we also expect the development of dusty outflows along the polar axis, which may naturally account for the polar dust emission recently detected in several AGNs from mid-infrared observations. Radiative feedback thus appears to be the key mechanism regulating the obscuration properties of AGNs, and we discuss its physical implications in the context of co-evolution scenarios. \\
\end{abstract}



\section{Introduction}

Active galactic nucleus (AGN) feedback can be driven by different physical mechanisms, such as jets, winds, and radiation pressure \citep[][and references therein]{Fabian_2012}. Radiation is the most direct way of communicating with the surrounding environment, and the matter-radiation coupling can be considerably enhanced if the interaction relies on radiation pressure on dust rather than electron scattering \citep{Fabian_1999, Murray_et_2005}. We recall that the standard Eddington luminosity for completely ionised gas is given by: $L_E = \frac{4 \pi G c}{\kappa_T} M_{BH}$, where $\kappa_T = \sigma_T/m_p$ is the electron scattering opacity, and $\sigma_T$ is the Thomson cross section. This is modified in the presence of dust, and the concept of `effective Eddington limit' for dusty gas has been introduced \citep{Fabian_et_2006}. As the dust absorption cross section is much larger than the Thomson cross section ($\sigma_d \gg \sigma_T$), the corresponding effective Eddington luminosity is much lower, and thus can be more easily exceeded. 

The relation between the boost factor ($A = \sigma_d/\sigma_T$), casted in terms of the classical Eddington ratio ($\lambda = L/L_E$), and the column density defines the $N_{\rm H} - \lambda$ plane \citep[e.g. see Figure 2 in][]{Fabian_et_2008}. In this plane, long-lived clouds can survive below the effective Eddington limit (to the left of the dividing line), while clouds in the forbidden region (to the right of the dividing line) see the central nucleus as above the effective Eddington limit  and should thus be outflowing. 
In this picture, the Compton-thick objects ($N_{\rm H} \gtrsim 10^{24} \rm cm^{-2}$) are always long-lived, provided that the Eddington ratio is lower than unity. We note that the trapping of reprocessed radiation has been assumed to be negligible in previous works \citep{Fabian_et_2006, Fabian_et_2008}. 

The $N_{\rm H} - \lambda$ plane has been observationally probed by plotting different samples of AGNs with measured $N_{\rm H}$ and $\lambda$ values, derived from X-ray observations. In general, the objects are found to respect the effective Eddington limit, by mostly avoiding the forbidden region \citep{Fabian_et_2008, Fabian_et_2009}. This has been confirmed by a deeper sample of AGNs based on the 2Ms Chandra Deep Field North-South \citep{Raimundo_et_2010}, as well as a detailed analysis of peculiar sources \citep{Vasudevan_et_2013}. Most recently, results based on the large and unbiased {\it Swift}/BAT AGN sample indicate that most sources keep away from the forbidden region, suggesting that the obscuration properties are primarily determined by the Eddington ratio \citep{Ricci_et_Nature}. This provides further confirmation that radiation pressure on dust is indeed the main mechanism regulating the obscuration properties of AGNs. 

Here we wish to revisit the $N_{\rm H} - \lambda$ plane from a theoretical perspective, and investigate how the inclusion of radiation trapping may modify the effective Eddington limit. 
We have previously shown that the dynamics and energetics of AGN-driven outflows, in particular the high momentum ratios and energy ratios observed in galactic outflows, can be reproduced by properly taking into account the effects of radiation trapping \citep{Ishibashi_Fabian_2015, Ishibashi_et_2018}. 
We have also discussed how AGN radiative feedback is capable of efficiently removing the obscuring dusty gas, and thus provide a natural physical interpretation for the observed AGN-starburst co-evolutionary sequence \citep{Ishibashi_Fabian_2016b}. We now examine how radiation trapping may affect the location of the forbidden region in the $N_{\rm H} - \lambda$ plane. 
We do not consider any specific geometry for the internal structure of the AGN (such as the putative torus in AGN unification scenarios), and assume a minimal configuration, consisting of the central black hole and its accretion disc, within a spherically symmetric gas distribution. We only discuss the effects of the anisotropic emission from the accretion disc, potentially leading to polar dusty outflows, in Section \ref{Anisotropic_radiation}.


\section{The $N_H - \lambda$ plane, with radiation trapping}

Dusty gas surrounding the AGN absorbs the ultraviolet (UV) radiation, thereby obscuring the central source, and re-emits in the infrared (IR) band. Here we explicitly consider the effects of radiation trapping, which has been neglected in previous works \citep{Fabian_et_2006, Fabian_et_2008}.


\subsection{The effective Eddington limit}

We assume that AGN radiation pressure sweeps up the ambient dusty gas into a shell with column density $N$. 
The radiative force is given by: 
\begin{equation}
F_{rad} = \frac{L}{c} \left(1 + \tau_{IR} - e^{-\tau_{UV}} \right) \, ,
\end{equation}
where $L$ is the central luminosity, $\tau_\mathrm{IR} = \kappa_\mathrm{IR} m_p N$ and $\tau_\mathrm{UV} = \kappa_\mathrm{UV} m_p N$ are the IR and UV optical depths, and $\kappa_{IR}$ and $\kappa_{UV}$ are the IR and UV opacities. 
The fiducial dust opacities are chosen such that $\kappa_{IR}/\kappa_T = 10 $ and $\kappa_{UV}/\kappa_T = 500$, corresponding to values of $\kappa_{IR} = 4 \, \mathrm{cm^2 g^{-1} f_{dg,MW}}$ and $\kappa_{UV} = 200 \, \mathrm{cm^2 g^{-1} f_{dg,MW}}$ (with the dust-to-gas ratio normalized to the Milky Way value). 

The gravitational force is given by:
\begin{equation}
F_{grav} = 4 \pi G m_p M_{BH} N \, , 
\end{equation} 
where the black hole mass ($M_{BH}$) is assumed to dominate the local gravitational potential. 
We recall that three distinct physical regimes can be identified according to the optical depth of the medium: optically thick to both IR and UV, optically thick to UV but optically thin to IR (single scattering limit), and optically thin to UV \citep[cf.][]{Ishibashi_Fabian_2015, Ishibashi_Fabian_2016b}. 

By equating the outward force due to radiation pressure and the inward force due to gravity, we obtain the effective Eddington luminosity:
\begin{equation}
L_E^{'} = \frac{4 \pi G c m_p M_{BH} N}{1 + \tau_{IR} - e^{-\tau_{UV}} } \, . 
\end{equation}  
The corresponding effective Eddington ratio is given by:
\begin{equation}
\Lambda = \frac{L}{L_{E}^{'} } = \frac{L (1 + \tau_{IR} - e^{-\tau_{UV}} )}{4 \pi G c m_p M_{BH} N}
\end{equation}  
Introducing the standard Eddington luminosity, $L_E$, and the classical Eddington ratio, $\lambda = L/L_E$, we obtain:
\begin{equation}
\Lambda = \frac{(1 + \tau_{IR} - e^{-\tau_{UV}}) \lambda}{\sigma_T N} \, . 
\end{equation} 
The effective Eddington limit is set by $\Lambda = 1$, which leads to the following condition on the column density:
\begin{equation}
N_E = \frac{(1 + \tau_{IR} - e^{-\tau_{UV}}) }{\sigma_T} \lambda \, . 
\label{Eq_N_E}
\end{equation}
This defines a critical column density ($N_E$), below which the dusty gas may be outflowing (see Section \ref{Section_blowout}).
In the IR-optically thick regime, Eq. (\ref{Eq_N_E}) can be approximated as
\begin{equation}
N_E \sim \frac{\lambda/\sigma_T}{(1 - \frac{\kappa_{IR}}{\kappa_T} \lambda) } \, . 
\label{Eq_N_E_IR}
\end{equation}


\subsection{The revised $N_{\rm H} - \lambda$ plane}

\begin{figure}
\centering
\begin{center}
\includegraphics[angle=0,width=0.4\textwidth]{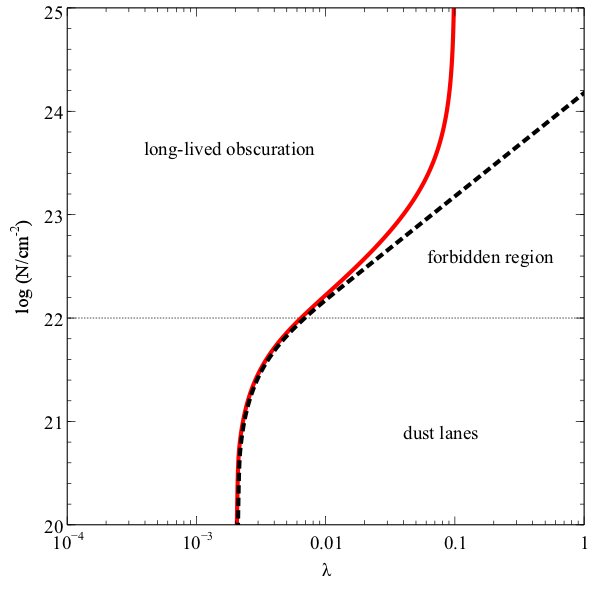}
\caption{\small
The revised $N_{\rm H} - \lambda$ plane: with radiation trapping (red solid) and single scattering-UV limit (black dotted). The horizontal line corresponds to $N = 10^{22} \rm cm^{-2}$ (grey fine-dotted).  
}
\label{plot_N_lambda_01}
\end{center}
\end{figure}

In Figure \ref{plot_N_lambda_01}, we plot the effective Eddington limit in the revised column density-Eddington ratio plane. The horizontal grey fine-dotted line marks the $N = 10^{22} \mathrm{cm}^{-2}$ limit, below which absorption by outer dust lanes may become important \citep{Fabian_et_2008, Fabian_et_2009}. 
We immediately observe that the forbidden region increases, with the inclusion of radiation trapping. 
In the present paper, we focus on the upper part of the $N_{\rm H} - \lambda$ plane (high column densities, $\log N \gtrsim 22$), as we are mainly interested in dealing with the effects of radiation trapping. 
We see that in the previously adopted single scattering regime ($\tau_{IR} \ll 1$), the critical column density increases linearly with the Eddington ratio, as $N_{E} \sim \frac{1}{\sigma_T} \lambda$ (black dotted diagonal line). In contrast, in the IR-optically thick regime ($\tau_{IR} \gg 1$), the effective Eddington ratio becomes independent of the column density (red solid curve). The two curves start to diverge beyond $\log N > 22$, with the difference becoming more apparent at higher column densities (where the IR term dominates). This implies that even dense material can potentially be disrupted in the IR-optically thick regime. 

Comparing the single scattering and the IR-optically thick regimes, we observe that in the latter case, a given Eddington ratio $\lambda$ can eject a higher column $N$. Conversely, a given column density $N$ can be ejected by a lower $\lambda$. In particular, with the inclusion of radiation trapping, even Compton-thick columns ($\log N \gtrsim 24$) can potentially be disrupted for Eddington ratios lower than unity. This is in stark contrast with the previously considered single scattering limit, in which Compton-thick clouds never see the nucleus as above the effective Eddington limit (provided that $\lambda < 1$), and are thus always long-lived. 
Therefore, by including radiation trapping, the shape of the forbidden region in the $N_{\rm H} - \lambda$ plane is significantly modified, and its area increases toward higher column densities and higher Eddington ratios. This suggests that previously long-lived clouds at high $N$ and high $\lambda$, are now located in the forbidden region and should thus be outflowing.


\subsection{The degree of radiation trapping}

We note that radiation trapping does not have to always happen. Initially, the gas is dumped into the nuclear regions in a quasi-spherical configuration, and the radiation pressure-driven shell is assumed to be spherically symmetric. As it expands, the outflowing shell may be disrupted and fragmented, giving rise to multiple clumps or clouds. The common occurrence of clouds being clumpy may be the result of the action of radiation pressure from the AGN or massive young stars. 

\begin{figure}
\centering
\begin{center}
\includegraphics[angle=0,width=0.4\textwidth]{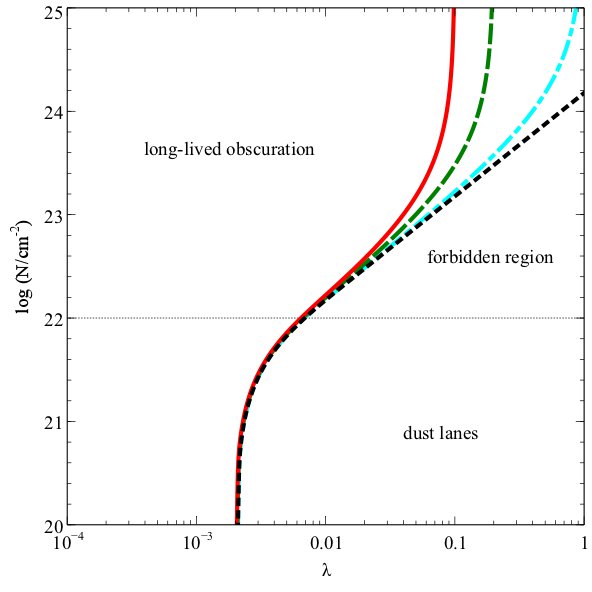} 
\caption{\small
The revised $N_{\rm H} - \lambda$ plane, including partial radiation trapping ($\tau_\mathrm{IR,eff} = C \tau_\mathrm{IR}$ with $C \leq 1$): maximal trapping (red solid), $C = 0.5$ (green dashed), $C = 0.1$ (cyan dash-dot), single scattering limit (black dotted). }
\label{plot_N_lambda_partial}
\end{center}
\end{figure}

Within cases when there is no complete shell, the reprocessed radiation tends to leak out through lower density channels, leading to a reduction in the effective optical depth. For instance, radiative transfer calculations indicate that the radiation force on dusty gas may be reduced by a factor of $\sim 2$ in cases of severe clumping compared to the case of a smooth gas distribution \citep{Roth_et_2012}.
If only partial trapping occurs, the effective $\tau_\mathrm{IR,eff}$ is lower, and the effective Eddington limit is shifted accordingly in the $N_{\rm H} - \lambda$ plane (Fig. \ref{plot_N_lambda_partial}). In the limit of negligible radiation trapping, we simply recover the single scattering limit. In fact, all possible cases of partial radiation trapping are bracketed between the maximal trapping (red solid curve) and the single scattering limit (black dotted curve).

\subsection{The impact of the dust-to-gas ratio}

\begin{figure}
\centering
\begin{center}
\includegraphics[angle=0,width=0.4\textwidth]{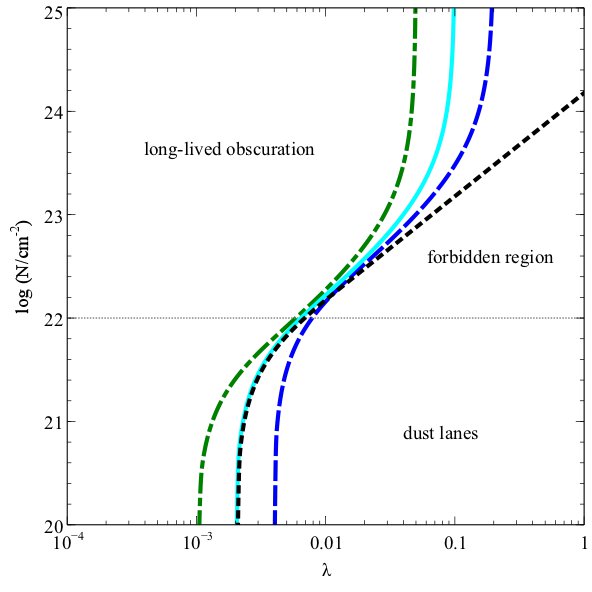}
\caption{\small
Same as Fig. \ref{plot_N_lambda_01}, with variations in the dust-to-gas ratio: $f_{dg} = 1 \times f_{\rm dg,MW}$ (cyan solid), $f_{dg} = 0.5 \times f_{\rm dg,MW}$ (blue dashed), $f_{dg} = 2 \times f_{\rm dg,MW}$ (green dash-dot).  
}
\label{plot_N_lambda_02}
\end{center}
\end{figure} 

We further consider the dependence of the effective Eddington limit on the underlying dust-to-gas ratio, $f_{dg}$.
Figure \ref{plot_N_lambda_02} shows the $N_{\rm H} - \lambda$ plane, with three different values of the dust-to-gas ratio, corresponding to $\times 1/2$, $\times 1$, and $\times 2$ times the Milky Way value ($f_{\rm dg,MW}$). 
We see that the three curves overlap in the single scattering regime (around $\log N \gtrsim 22$), where the material is optically thick to UV but optically thin to IR, as the effective Eddington ratio is independent of $f_{dg}$ in the single scattering limit. On the other hand, the curves increasingly diverge in the IR-optically thick regime (and also in the UV-optically thin regime). 

In general, we observe that the area of the forbidden region increases for increasing dust-to-gas ratios. 
This implies that the more dusty material is unlikely to survive as long-lived clouds, and can be more easily ejected for lower Eddington ratios. Indeed, for a higher dust-to-gas ratio $f_{dg}$, a given column density $N$ can be ejected by a lower $\lambda$. Therefore even high-density columns can be easily disrupted by sub-Eddington luminosities, even more so if they are dust-rich, suggesting that dust-rich clouds cannot be too long-lived close to the central nucleus.


\section{Comparison with observations}

\begin{figure}
\centering
\begin{center}
\includegraphics[angle=0,width=0.4\textwidth]{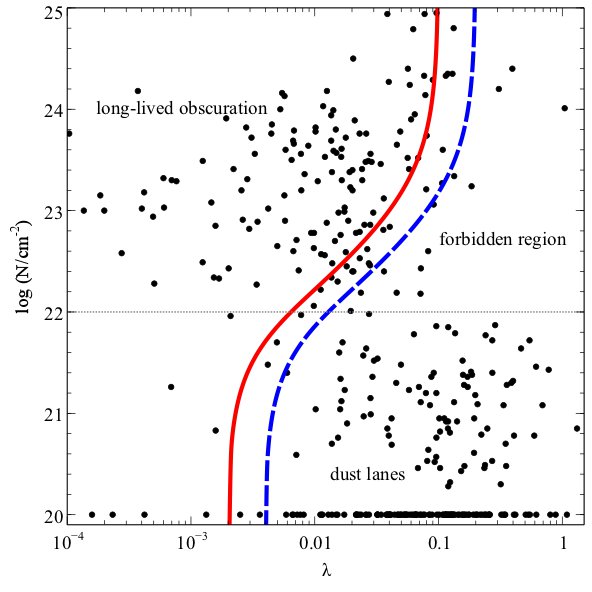} 
\caption{\small
The {\it Swift}/BAT AGN sample plotted on the revised $N_{\rm H} - \lambda$ plane: the red solid line indicates the effective Eddington limit including radiation trapping (for a standard Milky Way dust-to-gas ratio), the blue dashed line shows a factor of 2 increase in the effective Eddington limit (e.g. due to intervening stars). Unobscured sources are assigned an upper limit of $\log N = 20$.  
Observational data from \citet{Ricci_et_2015, Ricci_et_2017}. 
}
\label{plot_N_lambda_complete_obsdata_MBH}
\end{center}
\end{figure} 

Observationally, it has been shown that most sources in X-ray selected AGN samples respect the effective Eddington limit, by carefully avoiding the forbidden region in the $N_{\rm H} - \lambda$ plane \citep{Fabian_et_2008, Fabian_et_2009, Raimundo_et_2010, Ricci_et_Nature}. 
The recent analysis of the most complete sample of local AGNs with measured X-ray properties, based on the 70 month {\it Swift}/BAT AGN catalogue, indicates that the great majority of obscured sources (with $N_{\rm H} > 10^{22} \mathrm{cm}^{-2}$) are located below the effective Eddington limit, with only $\sim 1\%$ of the sample being found in the forbidden region \citep[see Figure 3 in][]{Ricci_et_Nature}. 

Here we plot the same {\it Swift}/BAT sample on the revised $N_{\rm H} - \lambda$ plane, including radiation trapping (Fig. \ref{plot_N_lambda_complete_obsdata_MBH}). 
We observe that the majority of objects still respect the modified effective Eddington limit, but a significant number of sources now lie in the enhanced forbidden region. Notably, a larger fraction of the sample (around $\sim 8\%$, with several borderline objects), is located in the extended forbidden region (delimited by the red solid line), the precise fraction also depending on the assumed dust-to-gas ratio.
 As discussed in \citet{Fabian_et_2009}, the gravitational force may be stronger if one includes the total enclosed mass (e.g. the mass of intervening stars). 
The blue dashed line in Figure \ref{plot_N_lambda_complete_obsdata_MBH} shows the effect of enclosing a mass equal to twice the central black hole mass, such that the effective Eddington limit is increased accordingly.  
This implies that a smaller fraction of the sample (around $ \sim 3\%$) is now located in the reduced forbidden region. 
The notable lack of sources in the forbidden region indicates that AGN radiation pressure plays a major role in regulating the amount of dusty gas in the host galaxy. This is unlikely to be due to selection effects, as the objects located in the forbidden region have higher Eddington ratios and hence should be brighter. 

In addition, we note that a handful of sources with Compton-thick columns ($\log N \gtrsim 24$) and Eddington ratios lower than unity ($\lambda < 1$), lie within the forbidden region in the revised $N_{\rm H} - \lambda$ plane. In principle, such sources see the central nucleus as exceeding the effective Eddington limit, and could potentially be in an outflowing phase (see also Discussion). 

Absorption by outer dust lanes located at larger radii in the host galaxy (on $\gtrsim$ kpc scales) cannot be too large, and the associated column densities cannot be too high, otherwise the required gas mass would be implausibly high. The choice of the lower limit at $N = 10^{22} \mathrm{cm^{-2}}$ appears to be consistent with the observational data, and is in line with Fig. 3 in \citet{Ricci_et_Nature}.


\section{Temporal evolution and blowout}
\label{Section_blowout}

Dusty gas is ejected when the central luminosity exceeds the effective Eddington limit. 
The forbidden region for long-lived clouds may thus correspond to a `blowout' region, where the objects should be experiencing outflows \citep[e.g.][]{Fabian_et_2008}. 
The temporal evolution of the obscuring column is governed by the balance between the radiative and gravitational forces. The corresponding equation of motion is given by:
\begin{equation}
\frac{d}{dt} [N r^2 v] = \frac{L}{4 \pi m_p c} (1 + \tau_{IR} - e^{-\tau_{UV}}) - G M_{BH} N \, . 
\end{equation} 
where $N$ is the shell column density, characterising the obscuration of the central source. 
Assuming that the gas follows an isothermal distribution \citep[as in][]{Fabian_et_2006}, the corresponding column density may be written as
\begin{equation}
N = \frac{N_0 R_0}{r} , 
\end{equation}
where $N_0$ is the initial column density, and $R_0 = 3$pc is the initial radius.
In this case, the column density falls off with radius as $\propto 1/r$, and so does the optical depth. In reality, the initial radius may not be strictly defined, e.g. in a clumpy distribution, and a lower limit is only set by the dust sublimation radius, below which electron scattering dominates the local opacity \citep[cf.][]{Ishibashi_Fabian_2015}. 
The effective Eddington ratio is given by
\begin{equation}
\Lambda = \frac{L r}{4 \pi G c m_p M_{BH} N_0 R_0} (1 + \tau_{IR} - e^{-\tau_{UV}}) \, . 
\end{equation}
Introducing again the classical Eddington ratio ($\lambda = L/L_E$), we obtain:
\begin{equation}
\Lambda =  \frac{r}{\sigma_T N_0 R_0}  (1 + \tau_{IR} - e^{-\tau_{UV}}) \lambda  \, . 
\label{Lambda_blowout}
\end{equation}
In the IR-optically thick regime, Eq. (\ref{Lambda_blowout}) can be rearranged as: 
\begin{equation}
\Lambda \sim ( \frac{r}{\sigma_T N_0 R_0} + \frac{\kappa_{IR}}{\kappa_T} ) \lambda \, . 
\end{equation}

\begin{figure}
\centering
\begin{center}
\includegraphics[angle=0,width=0.4\textwidth]{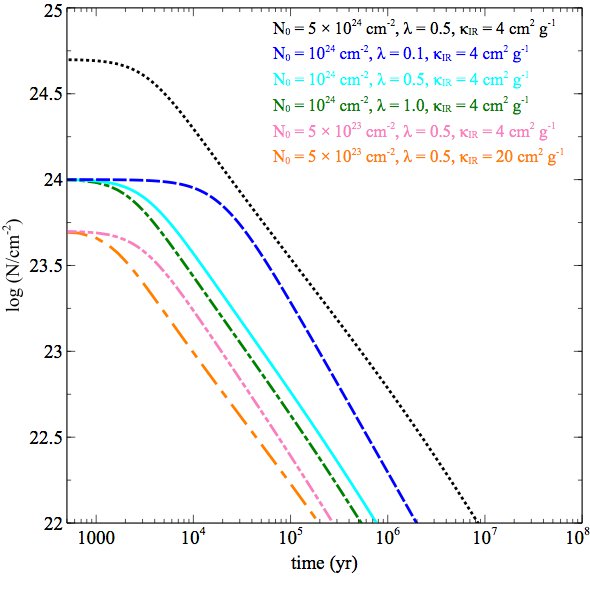} 
\caption{\small Temporal evolution of the obscuring column density: 
$N_0 = 5 \times 10^{24} \, \rm cm^{-2}$, $\lambda = 0.5$, $\kappa_{IR} = 4 \, \rm cm^2 g^{-1}$ (black dotted);
$N_0 = 10^{24} \, \rm cm^{-2}$, $\lambda = 0.1$, $\kappa_{IR} = 4 \, \rm cm^2 g^{-1}$ (blue dashed);
$N_0 = 10^{24} \, \rm cm^{-2}$, $\lambda = 0.5$, $\kappa_{IR} = 4 \, \rm cm^2 g^{-1}$ (cyan solid);
$N_0 = 10^{24} \, \rm cm^{-2}$, $\lambda = 1.0$, $\kappa_{IR} = 4 \, \rm cm^2 g^{-1}$ (green dash-dot);
$N_0 = 5 \times 10^{23} \, \rm cm^{-2}$, $\lambda = 0.5$, $\kappa_{IR} = 4 \, \rm cm^2 g^{-1}$ (pink dash-dot-dot);
$N_0 = 5 \times 10^{23} \, \rm cm^{-2}$, $\lambda = 0.5$, $\kappa_{IR} = 20 \, \rm cm^2 g^{-1}$ (orange dash-dot-fine).
}
\label{plot_N_t_var}
\end{center}
\end{figure}

Figure \ref{plot_N_t_var} shows the temporal evolution of the obscuring column for different values of the initial column density ($N_0$), Eddington ratio ($\lambda$), and dust-to-gas ratio ($f_{dg}$). 
For a given initial column density ($N_0 = 10^{24} \rm cm^{-2}$, corresponding to the Compton-thick limit), the obscuration falls off more rapidly for higher Eddington ratios, as can be seen by comparing the blue dashed ($\lambda = 0.1$), cyan solid ($\lambda = 0.5$), and green dash-dot ($\lambda = 1.0$) curves. 
On the other hand, for a given Eddington ratio ($\lambda = 0.5$), a higher initial column density naturally implies a higher obscuration with a slower dissipation, as seen by comparing the black dotted ($N_0 = 5 \times 10^{24} \rm cm^{-2}$), cyan solid ($N_0 = 10^{24} \rm cm^{-2}$), and pink dash-dot-dot ($N_0 = 5 \times 10^{23} \rm cm^{-2}$) curves. 
Finally, for the same initial conditions on $N_0$ and $\lambda$, an increase in the dust-to-gas ratio, and hence IR opacity ($\kappa_{IR} \propto f_{dg}$), leads to a faster dissipation of the obscuring material (pink dash-dot-dot vs. orange dash-dot-fine curves). 

We may define a characteristic blowout timescale, $t_{out}$, as the time required for the column density to drop to $\log N = 22$, where the source may be considered as unobscured. From Figure \ref{plot_N_t_var}, we see that even initially Compton thick material can be driven out at sub-Eddington luminosities (for $\lambda \gtrsim 0.1$), although it may take a rather long time for the obscuration to dissipate ($t_{out} \lesssim 10^7$yr). The blowout timescale decreases with increasing Eddington ratio, such that $t_{out} \sim10^6$yr for $\lambda \sim 0.5$. 
Of course, for a lower initial column density, the blowout timescale will be shorter ($t_{out} \gtrsim 10^5$yr). 
Overall, the outflowing timescale may be typically of the order of $\sim$million years, suggesting a rather short-lived blowout phase in the evolutionary sequence.   

The low percentage of objects in the forbidden region and the million-yr timescale required to push the gas out of the way have implications for large-scale luminosity variations of luminous obscured AGN. The paucity of objects in the forbidden region means that high-$\lambda$ episodes must be relatively long-lived, lasting of order a million years.

\section{Anisotropic radiation and polar outflows}
\label{Anisotropic_radiation}

In realistic situations, we should also take into account the effects of anisotropic radiation. The UV/optical emission is generally assumed to originate from the accretion disc: the emerging flux is maximal along the polar axis, and declines with increasing inclination angle. 
As a consequence, the effective Eddington ratio will also be a function of the inclination angle.

\begin{figure}
\centering
\begin{center}
\includegraphics[angle=0,width=0.4\textwidth]{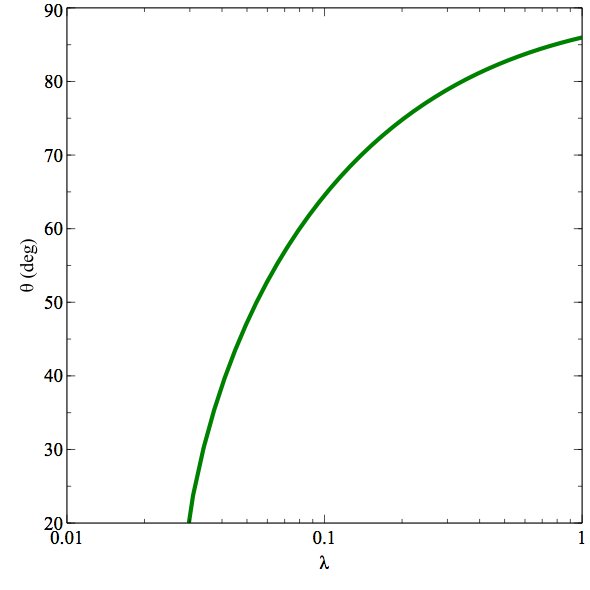} 
\caption{\small Opening angle ($\theta$) as a function of the Eddington ratio ($\lambda$) in the IR-optically thick regime. 
Note that the measured Eddington ratio is not the `true' Eddington ratio, due to the inclination angle (see main text). 
}
\label{plot_theta_lambda}
\end{center}
\end{figure} 

We may assume that the radiation flux follows a relation of the form: $F_{rad} \propto  \cos \theta (1+ 2 \cos \theta)$, which takes into account the change in the projected surface area and the limb darkening effect \citep[][and references therein]{Kawaguchi_Mori_2010, Liu_Zhang_2011}. 
The effective Eddington ratio, including the angular dependence, is then given by:
\begin{equation}
\Lambda \sim \frac{\cos \theta (1 + 2 \cos \theta) (1+ \kappa_{IR} m_p N)}{\sigma_T N} \lambda \, . 
\end{equation}
As a result, the effective Eddington ratio will be maximal along the polar axis, and will drop below unity at some critical angle. 
Setting $\Lambda = 1$ and solving for $\cos \theta$, we obtain:
\begin{equation}
\cos \theta = \frac{-1 + \sqrt{1+ 8  \frac{\sigma_T N}{(1 + \kappa_{IR} m_p N)} \frac{1}{\lambda}}}{4} \, . 
\end{equation} 
In the IR-optically thick limit, the above expression can be approximated as: 
\begin{equation}
\cos \theta = \frac{-1 + \sqrt{1+ 8 \frac{\kappa_T}{\kappa_{IR}} \frac{1}{\lambda}}}{4} \, .
\label{Eq_costheta}
\end{equation}  
This defines a critical opening angle of the outflow cone as a function of the Eddington ratio $\lambda$.  
In Figure \ref{plot_theta_lambda}, we plot the opening angle $\theta$ as a function of the Eddington ratio $\lambda$. The condition $\cos \theta \leq 1$ implies a minimum Eddington ratio, given by $\lambda \geq \frac{\kappa_T}{3 \kappa_{IR}}$. 
The opening angle increases with increasing Eddington ratio, for instance: $\theta \sim 65^{\circ}$ for $\lambda = 0.1$, and $\theta \sim 86^{\circ}$ for $\lambda = 1$. 
Such dusty outflows along the polar axis may provide a natural physical interpretation for the polar dust emission recently detected in several AGNs. 

Indeed, somewhat surprisingly, recent interferometric observations indicate that the bulk of the mid-IR emission arises from the polar region rather than from the classical `dusty torus' \citep{Hoenig_et_2013, Asmus_et_2016, Lopez-Gonzaga_et_2016}. In a number of nearby Seyferts, the polar dust emission is observed to extend on scales from a few to hundreds of parsecs, and is consistent with emission originating from a `dusty wind' in a hollow cone structure \citep[e.g.][and references therein]{Hoenig_Kishimoto_2017}. 
In contrast to the standard AGN unification schemes \citep{Antonucci_1993}, the new interferometric observations support a scenario in which the IR emission mainly originates from a dusty wind in the polar direction.
In our picture, dusty outflows along the polar axis are naturally expected, due to the anisotropic emission from the accretion disc, and we provide predictions as to the opening angle of the outflowing cone as a function of the Eddington ratio (Eq. \ref{Eq_costheta}). 

Within the above simple model for the disc emission, the measured Eddington ratio is not the angle-averaged Eddington ratio, due to the disc inclination. (We measure $\cos \theta (1+2\cos \theta) \cdot \lambda_0$ where $\lambda_0$ is the pole-on value.) Consequently most AGN in the forbidden or blowout region have an inclination which is less than the opening angle shown in Fig. \ref{plot_theta_lambda}. Objects close to, but to the right of the effective Eddington limit shown in Fig. \ref{plot_N_lambda_complete_obsdata_MBH} will have an inclination low enough that we view them within the opening angle shown in Fig. \ref{plot_theta_lambda}.

\begin{figure}
\centering
\begin{center}
\includegraphics[angle=0,width=0.4\textwidth]{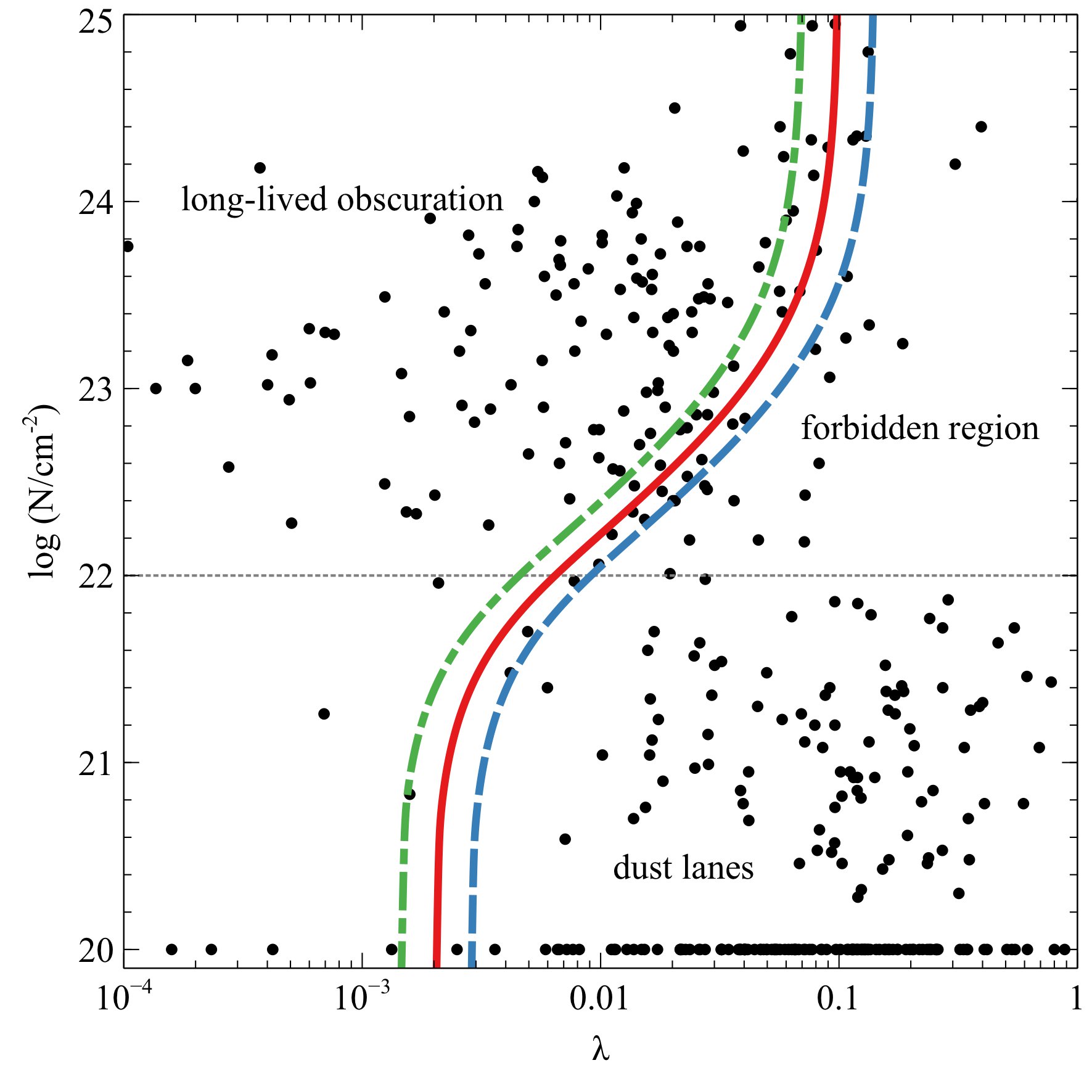} 
\caption{\small The revised $N_{\rm H} - \lambda$ plane for different inclination angles: $\theta = 0^{\circ}$ (green dash-dot), $\theta = 45^{\circ}$ (red solid), $\theta = 60^{\circ}$ (blue dashed). }
\label{plot_NH_lambda_45}
\end{center}
\end{figure} 

We note that dusty gas could be outflowing in one direction, but not in others. As the observational data are obtained for a given line of sight, the location of the sources on the $N_{\rm H} - \lambda$ plane (either in the forbidden or long-lived region) corresponds to that particular sightline. As an illustration, Fig. \ref{plot_NH_lambda_45} shows the effective Eddington limit for different inclination angles, assuming a simple $\propto$$\cos \theta$ dependence. We observe that a given source can be located in the forbidden region for low inclination angles, with gas outflowing along the polar direction; while the same system may be located in the long-lived region for higher inclination angles, as dusty gas survives in the edge-on direction. The outflowing gas may not be observable through absorption spectroscopy, but could be detected by emission spectroscopy in the polar directions. 

However, the above mentioned anisotropy effects need to be interpreted with caution. In reality, the radiation pattern is determined by relativistic effects, which become particularly important for rapidly spinning black holes. 
In qualitative terms, the relativistic effects such as gravitational focusing tend to redirect the radiation back to the disc, thus effectively reducing the anisotropy effect \citep{Sun_Malkan_1989, Li_et_2005}. Moreover, multi-dimensional and non-linear effects come into play in the IR-optically thick regime, such that the global evolution of radiation trapping can only be followed through 2D/3D numerical simulations \citep[e.g.][]{Roth_et_2012}.


\section{Discussion}

By explicitly taking into account radiation trapping, we obtain that the effective Eddington limit is modified, leading to an enhanced forbidden region in the $N_{\rm H} - \lambda$ plane (Fig. \ref{plot_N_lambda_01}). As a result, higher column densities can be ejected by lower Eddington ratios, and previously long-lived clouds may now be outflowing. This implies that even Compton-thick material can potentially be disrupted for sub-Eddington luminosities. Moreover, an increase in the dust-to-gas ratio leads to a further increase in the area of the forbidden region, such that the more dusty gas is more easily expelled (Fig. \ref{plot_N_lambda_02}). Therefore, Eddington-limited radiative feedback can be more than sufficient to clear out the obscuring dusty gas.
We emphasise the remarkable absence of sources in the forbidden region (Fig. \ref{plot_N_lambda_complete_obsdata_MBH}), which suggests that radiation pressure on dust is indeed the key mechanism coupling the AGN to the surrounding gas. 

Sources located in the forbidden region of the $N_{\rm H} - \lambda$ plane are likely caught in the outflowing phase, when the central nucleus is expelling its obscuring cocoon in a short-lived blowout event. In fact, the pre-selection of objects located in the forbidden region seems to be an efficient criterion to pick up AGN radiatively-driven outflows \citep{Kakkad_et_2016}. Indeed, powerful outflows have been detected in X-ray obscured QSOs at $z \sim 1.5$ \citep{Brusa_et_2015, Perna_et_2015}. Recent observations have also uncovered new populations of dust-reddened quasars, which are good candidates for sources `in transition', from the dust-obscured starburst phase to the unobscured luminous quasar stage \citep{Banerji_et_2012, Banerji_et_2015, Glikman_et_2012, Zakamska_et_2016}. A particularly high fraction of such red quasar population is found to show BAL characteristics  \citep{Urrutia_et_2009, Glikman_et_2012}, while some of the dusty quasars show direct evidence for outflows in the form of broadened $H\alpha$ line profiles \citep{Banerji_et_2012}. Estimates based on the observed space densities indicate that this emergence phase is short-lived, and typically lasts a few $\sim 10^6 $yr \citep{Glikman_et_2012}, comparable to the blowout timescale discussed in Section 4. Interestingly, a recent study based on the FIRST and 2MASS surveys suggests that dust-reddened quasars are located right in the forbidden region of the $N_{\rm H} - \lambda$ plane \citep{Glikman_2017}. We have previously analysed how our AGN radiative feedback-driven shell models may be applied to the different populations of dusty quasars reported in recent observations \citep{Ishibashi_et_2017}. 
Indeed, AGN radiative feedback may naturally account for related aspects, such as the development of large-scale dusty outflows, which are now starting to be observed \citep{Zakamska_et_2016, Hamann_et_2017}. 

The combination of high Eddington ratio and high dust content should form favourable conditions for AGN radiative feedback. At higher redshifts, near the peak epoch of both AGN and star formation activities ($z \gtrsim 2$), we can thus expect an increased number of sources close to or above the effective Eddington limit in the $N_{\rm H} - \lambda$ plane. In particular, the enhanced forbidden region (due to radiation trapping) may be more densely populated by high-density, dust-obscured, ULIRG-like systems at high redshifts. 
This can be observationally tested by analysing deeper samples at higher redshifts nowadays available. By carefully comparing the potential differences between AGN samples at low and high redshifts, we may gain new insight into their global evolution. For those sources located in the forbidden region, it will also be interesting to perform follow-up multi-wavelength observations in order to search for direct outflow signatures. 

In general, the obscuration properties of AGNs have been discussed in the framework of unification scenarios based on the `obscuring torus' and variants thereof \citep[e.g. since][]{Antonucci_1993}. A common view is that orientation effects may account for the different classes of AGNs (Type I vs. Type II), as supported by observations of polarised optical radiation. However, polarisation signatures are only detected in a limited number of sources, and at present there is no compelling evidence for a simple obscuring torus geometry, possibly suggesting a more complex picture \citep[see the recent review by][and references therein]{Netzer_2015}. 

The different AGN obscuration models discussed so far may be broadly divided into two categories: static and dynamical. 
For instance, analytic calculations suggest that the vertical support for a geometrically thick obscuring torus can be provided by infrared radiation pressure \citep{Krolik_2007}. The different nature of smooth and clumpy dust tori in different AGN luminosity regimes has been analysed by \citet{Hoenig_Beckert_2007}. 
Circulation of gas leading to the development of radiation-driven `fountains' has also been proposed as a possible mechanism for the formation of the obscuring torus \citep{Wada_2012}.  
The detailed evolution of dusty gas clouds illuminated by the central AGN has been followed through 3D radiation hydrodynamic (RHD) simulations \citep{Namekata_et_2014}; while the dynamics of dusty gas subject to anisotropic disc emission, X-ray heating, and reprocessed IR radiation, has also been analysed via axisymmetric RHD simulations \citep{Namekata_Umemura_2016}. 

Another class of models consider the obscuring toroidal structure as supported by winds driven by IR radiation pressure, due to the relatively high opacity of dust in the infrared \citep{Dorodnitsyn_et_2011}. 
A further refinement of the model based on multi-dimensional RHD simulations confirms that the geometrically thick obscuration on pc-scales can indeed be provided by outflows driven by IR radiation pressure on dust \citep{Dorodnitsyn_et_2012}. 
On the other hand, 3D time-dependent RHD simulations of dusty tori suggest that UV radiation pressure on dust can drive powerful winds with considerable mass loss rates \citep{Chan_Krolik_2016}. The most recent multi-dimensional simulations of dusty gas acceleration in star-forming galaxies suggest that dusty clouds are able to survive longer when accelerated by radiation pressure rather than entrained in a hot outflow \citep{Zhang_et_2018}.

Here we envisage a simple picture in which AGN radiative feedback, driven by radiation pressure on dust, strongly affects the distribution of dusty gas in the host galaxy. We assume a spherical shell geometry throughout the paper, while we separately discuss the case of anisotropic emission from the accretion disc in Section \ref{Anisotropic_radiation}. Our results are in agreement with the observational data based on the 70 month {\it Swift}/BAT AGN catalogue, which is the most complete sample of X-ray selected AGNs, with measured $N_H$ and $\lambda$ values, currently available. This suggests that our model assumptions are reasonable and sufficient for the purpose of comparison with X-ray observations.

In summary, the overall picture is entirely consistent with AGN radiation pressure on dust being the main physical mechanism regulating the co-evolutionary sequence, linking dust-obscured starbursts to unobscured luminous quasars \citep{Ishibashi_Fabian_2016b}. The global evolution may be governed by a complex interplay between accretion luminosity and dusty gas obscuration. At early times, following a merger event or a more secular process, large amounts of gas are funnelled toward the centre, efficiently feeding the accreting black hole, but also providing significant obscuration. A positive correlation between high Eddington ratio and high column density is thus observed in this early stage. If the system enters in the IR-optically thick regime (characterised by efficient radiation trapping), the obscuring dusty gas may be ejected by AGN radiative feedback. Higher Eddington ratios imply stronger outflows, which rapidly clear out the obscuring cocoon, leading to an anti-correlation between luminosity and obscuration. The central source may briefly shine as a bright quasar, until the accretion disc is drained. Following the removal of dusty gas, the AGN may decline in luminosity, giving rise to unabsorbed sources at low Eddington ratios.


\section*{Acknowledgements }

WI acknowledges support from the University of Zurich. ACF and WI acknowledge ERC Advanced Grant 340442.
CR acknowledges financial support from the China-CONICYT fund, the CONICYT-Chile grants FONDECYT 1141218, and Basal-CATA PFB--06/2007.

  
\bibliographystyle{mn2e}
\bibliography{biblio.bib}

\label{lastpage}

\end{document}